\documentclass[%
 reprint,
superscriptaddress,
nofootinbib,
 amsmath,amssymb,
 aps,eqsecnum,
]{revtex4-2}

\usepackage{graphicx}
\usepackage{dcolumn}
\usepackage{bm}

\usepackage{graphicx}
\usepackage[dvipsnames, usenames]{xcolor}

\definecolor{linkcolor}{rgb}{0.0,0.3,0.5}
\usepackage[colorlinks=true,allcolors=linkcolor]{hyperref}
\usepackage[normalem]{ulem} 
\usepackage{float}

\newcommand{\sid}[1]{\textcolor{magenta}{\textbf{#1}}}

\begin{document}

\preprint{APS/123-QED}

\title{Spacetime Quasi-normal Mode Oscillations of Anisotropic Neutron Stars}

    \author{Jihao Yu}
        \affiliation{Department of Physics, University of Virginia, Charlottesville, VA 22904, USA}

    \author{Victor Guedes}
        \affiliation{Department of Physics, University of Virginia, Charlottesville, VA 22904, USA}

            \author{Shu Yan Lau}
    \affiliation{Department of Physics, Montana State University, Bozeman, MT 59717, USA}

    \author{Siddarth Ajith}
    \affiliation{Department of Physics, University of Virginia, Charlottesville, VA 22904, USA}

    \author{Kent Yagi}
    \affiliation{Department of Physics, University of Virginia, Charlottesville, VA 22904, USA}
    
\date{\today}

\begin{abstract}
Neutron star asteroseismology offers a unique opportunity to probe nuclear physics through stellar oscillations. Although the pressure inside of a neutron star is typically assumed to be isotropic, pressure anisotropy can arise from various physical mechanisms, including elasticity, viscosity, and magnetic fields. Previous studies of nonradial stellar quasi-normal mode oscillations with anisotropic pressure have focused primarily on fluid modes. In this paper, we compute, for the first time, spacetime oscillation modes (so-called $w$-modes) of anisotropic neutron stars. Using a perturbative framework for stellar oscillations with pressure anisotropy, developed previously by some of the authors, together with a phenomenological anisotropy model, we find that both the real and imaginary parts of the $w$-mode frequencies decrease as the tangential pressure becomes dominant over the radial pressure. Although we do not find any unstable $w$-modes within the physically viable parameter space, unstable $w$-modes appear in an unphysical branch of solutions when the tangential pressure strongly dominates the radial one. We also find that the relation between the real part of the $w$-mode frequency and the stellar compactness is quasi-universal with respect to variations in the equation of state and the degree of pressure anisotropy. In contrast, the relation between the imaginary part of the $w$-mode frequency and the stellar compactness depends on the degree of anisotropy, but remains equation-of-state universal when the anisotropy is fixed. Finally, we discuss potential mode crossings and the validity of certain approximations that have been shown to work well for $w$-mode calculations in the isotropic case. 
\end{abstract}

\maketitle

\section{\label{sec:level1}Introduction
\protect}

Neutron stars (NSs) are extremely compact objects whose central density exceeds the nuclear saturation density and thus provide an ideal laboratory environment for probing nuclear physics. Various NS observations through radio~\cite{Antoniadis:2013pzd,NANOGrav:2019jur,Saffer:2024tlb}, x-rays~\cite{Raaijmakers:2021uju,Yunes:2022ldq}, gamma-rays~\cite{Guedes:2024zkh}, and gravitational waves~\cite{LIGOScientific:2018cki,Chatziioannou:2020pqz,Yunes:2022ldq} have placed constraints on the equation of state (EOS) for nuclear matter, the relation between energy density and pressure.

One useful approach for extracting nuclear physics information from NSs is stellar seismology~\cite{Andersson:1996pn,Andersson:1997rn,Kokkotas:1999bd}, the study of oscillations of astronomical objects. The $\ell > 1$ nonradial oscillation modes of NSs are quasi-normal modes, characterized by complex frequencies due to damping over time via gravitational wave emission.
These quasi-normal modes carry information about nuclear physics through their dependence on the interior properties of NSs, which are governed by the underlying EOS of nuclear matter. Quasi-normal modes can be classified based on the major restoring forces or properties of the composite waves, including the fundamental mode ($f$-mode), the pressure mode ($p$-mode), and the spacetime mode ($w$-mode)~\cite{Kokkotas:1999bd}.

Quasi-normal modes are also useful for studying the (non-radial) stability of stellar modes. Stellar oscillations are conventionally described by containing a time-dependent oscillation function, $e^{i\omega t}$, where $\omega$ is the frequency and $t$ is the time. A stellar oscillation mode is considered stable if the complex frequency of the quasi-normal mode has a positive imaginary part such that $e^{i\omega t}$ is exponentially decaying. Detweiler and Ipser~\cite{ipser1973} have developed a method that argues the stability of quasi-normal modes through a variational method. They found that all spherically symmetric equilibrium, isotropic NSs with a positive density, positive pressure, negative density gradient, and non-negative Schwarzschild discriminant have stable quasi-normal modes\footnote{Convective instability can still occur via a negative Schwarzschild discriminant.}.

Most studies on NSs focus on isotropic pressure, while some degree of anisotropy in pressure may exist inside NSs. Such anisotropy can arise from different origins, including viscosity~\cite{viscosity}, elasticity~\cite{elasticity,Karlovini:2002fc,Dong:2024lte}, superconductivity~\cite{HERRERA199753}, and strong magnetic field~\cite{Most_2025}. Pressure anisotropy also exists in  solutions to the Einstein equations for exotic compact objects, such as boson stars~\cite{Macedo:2013jja}, gravastars~\cite{Chirenti_2007,Cattoen_2005}, and dark energy stars~\cite{lopes}. 

There are several studies on non-radial oscillations for anisotropic NSs. These modes were first calculated within the Cowling approximation~\cite{Cowling:1941nqk}, where only fluid perturbations are kept while spacetime perturbations are ignored. It is only recently that the complete perturbative framework for computing non-radial oscillations for anisotropic NSs has been developed in full general relativity~\cite{nonradial} (see~\cite{Mondal:2023wwo,Mondal:2023wwo,Arbanil:2025jep} for related works). In this reference, some of us computed the $f$-modes and $p$-modes of NSs with some phenomenological anisotropy models and found that, unlike isotropic NSs, anisotropic NS $p$-modes can become unstable. They justified this numerical finding with some analytic calculations by extending the variation principle method by Detweiler and Ipser~\cite{ipser1973} to anisotropic NSs, proving that the modes can become unstable once the anisotropy is turned on. 
 
A recent work~\cite{Guedes:2025gqi} by some of us studied quasi-universal relations between the real part of the $f$-mode frequency and the tidal deformability for anisotropic NSs, which was known to exist for isotropic NSs~\cite{Chan:2014kua} (see e.g.~\cite{Yagi:2015hda} for another work on quasi-universal relations for anisotropic NSs). For the anisotropic case, we found that the relations depend strongly on the amount of anisotropy, while they remain EOS-insensitive for a fixed anisotropy. Using these relations, together with the inference on the $f$-mode frequency and the tidal deformability of the binary NS merger event GW170817~\cite{Pratten:2019sed}, we derived EOS-insensitive bounds on anisotropy. 

One type of oscillation mode for anisotropic NSs that has not been computed previously is the $w$-mode~\cite{1992MNRAS.255..119K,Andersson_1996}. $w$-modes are also known as spacetime modes since they originate from the metric perturbations, so they cannot be computed under the Cowling approximation. Thanks to the recent development of the perturbative framework for stellar oscillations with pressure anisotropy in full general relativity~\cite{nonradial}, we are now able to compute, for the first time, polar $w$-mode oscillations for anisotropic NSs, which is the goal of this paper. We solve the same perturbation equations in~\cite{nonradial} that we used to compute $f$-modes and $p$-modes. In contrast to the fluid modes ($f$-modes and $p$-modes, etc.) which have an imaginary part of the frequency that is orders of magnitude smaller than the real part, $w$-mode frequencies have comparable imaginary and real parts. Therefore, we use Leaver's continued fraction method~\cite{1985Leaver} to compute the $w$-mode frequencies.

\begin{figure}
    \centering
    \includegraphics[width=1\linewidth]{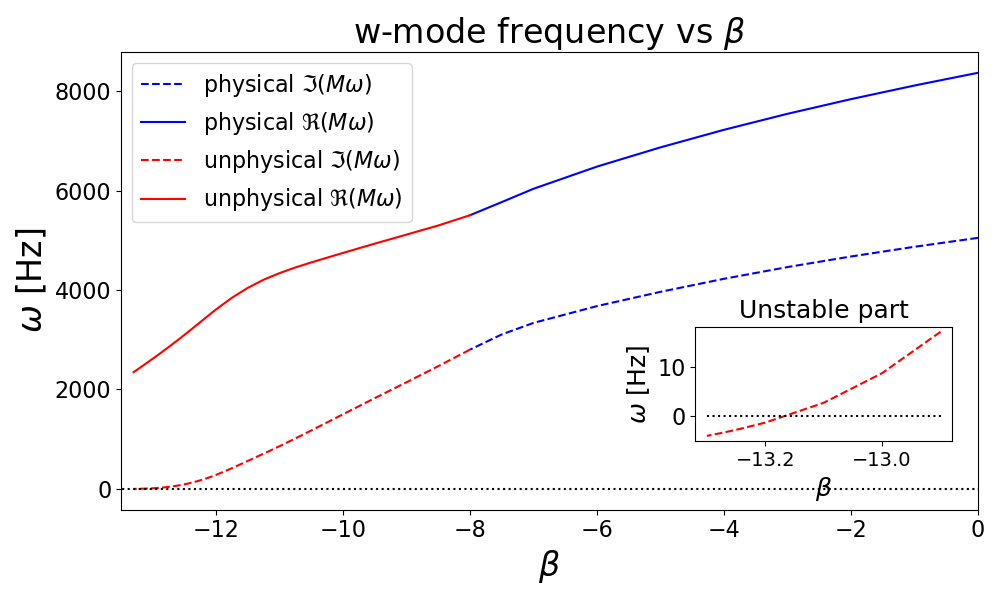}
    \caption{ Frequencies of $w$-modes against the anisotropy parameter $\beta$ for NSs with MS1 EOS  and central density of $\rho_c=6\times10^{14}$g/cm$^3$. We indicate physical branches as blue while unphysical branches are red,  where the tangential sound speed exceeds unity and thus violates causality. 
    (Subplot) Zoomed-in version for largely-negative $\beta$, where the imaginary part of the frequency becomes negative, indicating the presence of unstable $w$-modes.}
    \label{fig:showunstable}
\end{figure}

Figure~\ref{fig:showunstable} summarizes our main result, which presents the $w$-mode frequencies as a function of the anisotropy parameter $\beta$, defined in Eq.~\eqref{eq:anisotropy_model}, for a selected EOS and central density. The dimensionless anisotropy parameter $\beta$ denotes the extent of pressure difference between the radial and tangential directions, where $\beta = 0$ corresponds to the isotropic case while $\beta < 0$ means that the tangential pressure dominates the radial one. The blue branches are for physically viable models, while the red ones are for unphysical models that violate causality and positivity of the tangential pressure for this specific choice of parameters (other physicality conditions can be violated for other choices of parameters). 
First, observe that both the real and imaginary parts of the $w$-mode frequency decrease as we decrease the anisotropy parameter $\beta$. Second, although we did not identify any unstable modes within the physically-viable branch, $w$-modes can become unstable for unphysical models with largely negative values of the anisotropy parameter. The analytic argument in~\cite{nonradial} for potential mode instabilities for anisotropic NSs did not specify the type of modes, and hence, it applies to the unstable $w$-modes found here. 

Are there any quasi-universal relations for $w$-mode frequencies that are insensitive to EOSs? For isotropic NSs, both real and imaginary frequencies enjoy universal relations with the stellar compactness~\cite{Benhar:2004xg,TL,Tsui:2005zf}. For anisotropic NSs, we found that the relation between the real part of the $w$-mode frequency and compactness remain quasi-universal against variations in both EOSs and the anisotropy parameter $\beta$. On the other hand, the relation between the imaginary part of the $w$-mode frequency and compactness depends sensitively on $\beta$, while it remains EOS-insensitive for fixed $\beta$. The latter is similar to the relation between the $f$-mode frequency and tidal deformability mentioned earlier~\cite{Guedes:2025gqi}.

The rest of the paper is organized as follows. In Sec.\ref{sec:perturbation framework section} we review how to construct background solutions and the perturbative framework for computing stellar oscillations with pressure anisotropy. In Sec.~\ref{sec:numertical results}, we provide the phenomenological anisotropy model and present numerical results for both background solutions as well as $w$-mode frequencies. We conclude in Sec.~\ref{sec:conclusion} and give some discussions, including potential mode crossing and the validity of certain approximations for computing $w$-mode frequencies that are known to work well for isotropic NSs~\cite{Andersson_1996,Wu_2007}.

\section{Perturbation framework}
\label{sec:perturbation framework section}

In this section, we review the perturbative framework for computing $w$-mode oscillations for anisotropic NSs following~\cite{nonradial} (see~\cite{Mondal:2023wwo,Mondal:2025ixk,Arbanil:2025jep} for related works). We first describe how to construct a spherically-symmetric background configuration in $c=G=1$ units. The line element is given by
\begin{eqnarray}
    ds^2 = \bar g_{\mu\nu} dx^\mu dx^\nu,
\end{eqnarray}
with
\begin{equation}
    \bar{g}_{\mu \nu} = \left(\begin{array}{cccc}
       -e^{\nu(r)} &  0 & 0 & 0\\
        0 &  e^{\lambda(r)}&0&0\\
        0 &0&r^2&0\\
         0&0&0&r^2 \sin^2\theta \\
    \end{array}\right).
\end{equation}
The  stress energy tensor for the anisotropic fluid is given by~\cite{1974StressEnergyTensor} 
\begin{equation}
    T_{\mu\nu}=\rho u_\mu u_\nu+p_rh_{\mu\nu}-\sigma\Omega_{\mu\nu}.
\end{equation}
Here, $h_{\mu\nu}=g_{\mu\nu}+u_\mu u_\nu $ and $\Omega_{\mu\nu}=h_{\mu\nu}-k_\mu k_\nu$, where $\vec{u}$ is the four velocity of the fluid element, and $\vec{k}$ is the unit radial vector perpendicular to $\vec{u}$. $\rho$ and $p_r$ are the energy density and the radial pressure, while $\sigma = p_r - p_t$ is the pressure anisotropy between the radial and tangential directions. 

Now, we can solve the NS perturbation background variables. Plugging the above line element and the stress-energy tensor into the Einstein equations, one finds a set of modified Tolman-Oppenheimer-Volkoff (TOV) equations for anisotropic stars~\cite{Horvat_2011}:
\begin{align}
\label{eq:m_TOV}
    m'&=4\pi r^2\rho,\\
    \label{eq:nu_TOV}
    \nu'&=2\frac{m+4 \pi r^3 p_r}{r^2}e^\lambda,\\
    \label{eq:pr_TOV}
    p_r'&=-(\rho +p_r)\frac{\nu'}{2}-\frac{2\sigma}{r}.
\end{align}
Here, a prime denotes a radial derivative, and $m(r)$ is defined as $e^\lambda=\left(1-\frac{2m}{r}\right)^{-1}$ which corresponds to the mass enclosed within a sphere of radius $r$.

Let us now introduce a linear perturbation to both the gravity and matter sectors. The metric is perturbed as
\begin{equation}
    g_{\mu \nu} = \bar{g}_{\mu\nu}+\delta g_{\mu\nu}\,,
\end{equation}
with 
\begin{eqnarray}
    \delta g_{\mu \nu} &=& \left(\begin{array}{cccc}
       e^\nu(r) H_0(r)  &  i\omega r H_1(r) & 0 & 0\\
        i \omega rH_1(r) & e^\lambda H_2(r) &0&0\\
        0 &0&r^2K(r)&0\\
         0&0&0&r^2 \sin^2\theta K(r)\\ 
          
    \end{array}\right) \nonumber \\
    && r^\ell\, Y_{\ell m}(\theta,\phi) e^{i\omega t},
\end{eqnarray}
where $Y_{\ell m}$ are spherical harmonics in spherical coordinates and $\omega$ is the angular frequency of the perturbation.

To study perturbations to the anisotropic fluid, we introduce the fluid displacement vector $\vec \zeta$ as~\cite{nonradial}
\begin{align}
\zeta^r &= \frac{W(r)}{r e^{\lambda(r)/2}}Y_{\ell m}(\theta,\phi)\, e^{i\omega t},\\
\zeta^\theta &= \frac{V(r)}{r^2} \partial_\theta Y_{\ell m} (\theta,\phi)\,e^{i\omega t},\\
\zeta^\phi &=  \frac{V(r)}{r^2 \sin^2\theta} \partial_\phi Y_{\ell m} (\theta,\phi)\,e^{i\omega t}.\label{Eq:fluid displacement}
\end{align}
We then arrive at the perturbation equations from the perturbed Einstein equations,
\begin{equation}
    \delta G_{\mu\nu}=8\pi \delta T_{\mu\nu},
\end{equation}
and the perturbed equation of motion,
\begin{equation}
    \delta \nabla_\mu T^{\mu\nu}=0.
\end{equation}

The anisotropic perturbation equations are derived from the Einstein field equations and matter equations of motion as in \cite{nonradial} (see Appendix~\ref{app:pert}).
We follow the general method as  described in \cite{LD}. To solve the frequency of the oscillation mode, we first solve for the stellar background by integrating the modified TOV equation. After that, we observe that the above set of equations establish an initial value problem with the vector function
\begin{eqnarray}
\vec{Y}=(H_1(r), K(r), W(r), X(r)).\label{eq:eigen_vec}
\end{eqnarray}
For a given oscillation frequency $\omega$, we can solve the perturbation functions throughout the star. Finally, we use Leaver's continued fraction method~\cite{1985Leaver,Sotani:2001bb}. 

\section{Numerical Results}
\label{sec:numertical results}
In this section, we provide details of the procedure for our numerical calculations and their results.
We adopt the same anisotropy model as in~\cite{nonradial,Guedes:2025gqi} that is an extension of the model used in~\cite{Horvat_2011}. The pressure difference between the radial and the tangential directions is given by
\begin{align}
\label{eq:anisotropy_model}
    \sigma= \beta p_r \mu^2,
\end{align}
where $\mu = 2m/r$ and $\beta$ is the dimensionless anisotropy parameter.

For equations of state (EOSs) connecting the radial pressure and energy density, we use the following: WFF1, SLy4, and MS1. WFF1 (MS1) is a soft (stiff) EOS, while SLy4 is in between.

\subsection{Spherically-symmetric NSs with Pressure Anisotropy}
\label{sec:SS}

We begin by constructing anisotropic NSs with \emph{spherically-symmetric} configurations, which will be used as background solutions when considering stellar perturbations in the subsequent subsection.
To achieve this, we solve the anisotropic TOV equations in Eqs.~\eqref{eq:m_TOV} and~\eqref{eq:pr_TOV} numerically. The boundary conditions at $r=r_0 \ll R$ (where $R$ represents the stellar radius) near the stellar center 
are given by
\begin{align}
    p_r &= p_c + \mathcal{O}(r_0^2),\\
    \rho &= \rho_c + \mathcal{O}(r_0^2),\\
    m &= \frac{4\pi}{3} \rho_c r_0^3+ \mathcal{O}(r_0^5).
\end{align}

For a chosen central pressure $p_c$ or central energy density $\rho_c$, we use the above asymptotic behaviors to solve Eqs.~\eqref{eq:m_TOV} and~\eqref{eq:pr_TOV}.  
The radius is determined from the condition $p_r(R)=0$ while the stellar mass is defined as $M = m(R)$. Eq.~\eqref{eq:nu_TOV} for $\nu$ is then solved under the boundary condition 
\begin{equation}
    e^{\nu(R)} = 1-\frac{2M}{R}\,.
\end{equation}

\begin{figure}
    \centering
    \includegraphics[width=0.99\linewidth]{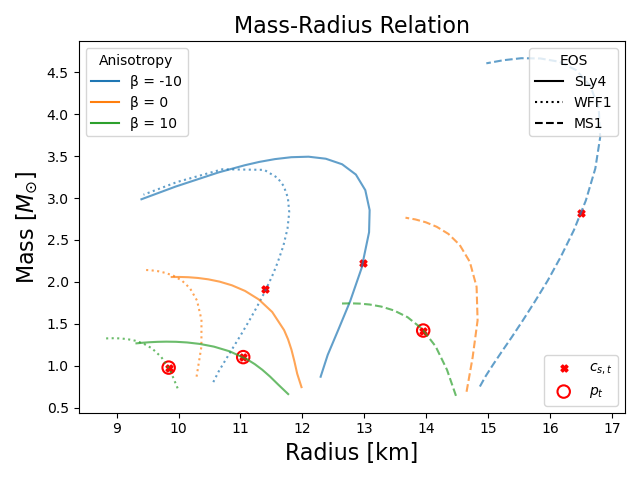}
    \caption{Mass-radius relations for anisotropic NSs. We present the relations for three different EOSs and anisotropy parameters. The red circles and crosses denote transition points from physical to unphysical configurations due to the positive pressure condition (denoted as $p_t$) and  causality (denoted as $c_{s,t}$)respectively. NSs with masses higher than the turning points are considered unphysical.}
    \label{fig:M-R relation}
\end{figure}

Figure~\ref{fig:M-R relation}
presents the mass-radius relation for anisotropic NSs. Observe that the NS radius and maximum mass increase (decrease) for negative (positive) $\beta$ from the isotropic case. Similarly to the isotropic case~\cite{Tassoul1978}, stars with the anisotropy model in Eq.~\eqref{eq:anisotropy_model} become unstable after reaching the maximum mass as one increases the central energy density~\cite{Guedes:2025gqi}. 

\begin{figure}[H]
    \centering
    \includegraphics[width=1\linewidth]{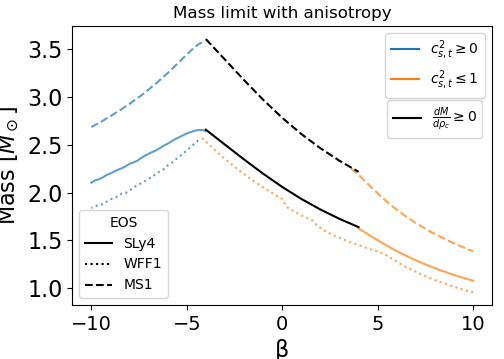}
    \caption{Maximum mass limit as a function of $\beta$ due to radial stability (black) and tangential causality (colored) for different EOSs. For a given anisotropy and EOS, the physical mass range is below the curves. For WFF1 EOS, the radial stability mass limit is always above the tangential causality limit for the range of central density we study, and therefore tangential causality is the only factor that sets the upper limit here.
    }
    \label{fig:placeholder}
\end{figure}

Below, we only consider stellar configurations that are physically viable, i.e. those that respect the energy conditions, positivity of pressure, and causality~\cite{Guedes:2025gqi}:

\begin{widetext}
    \begin{align}
                &\textrm{1. weak energy condition~\cite{Poisson:2009pwt}: } \quad \rho\geq0,\quad \rho+p_{r}>0,\quad \rho+p_{t}>0, \label{wec}\\
                &\textrm{2. null energy condition~\cite{Poisson:2009pwt}: }\quad \rho+p_{r}\geq0 \quad \rho+p_{t}\geq0, \label{nec}\\
                &\textrm{3. strong energy condition~\cite{Poisson:2009pwt}: }\quad \rho+p_{r}+2p_{t}\geq0, \quad \rho+p_{r}\geq0, \quad\rho+p_{t}\geq0, \label{sec}\\
                &\textrm{4. dominant energy condition~\cite{Poisson:2009pwt}: }\quad \rho\geq0, \quad \rho\geq|p_{r}|, \quad \rho\geq|p_{t}|, \label{dec} \\
                &\textrm{5. positivity of pressure: }\quad  p_{r}\geq0,  \quad p_{t}\geq0, \\
                &\textrm{6. causality: }\quad  0 \leq c^{2}_{s,r}, c^{2}_{s,t} \leq 1. \label{causality}
            \end{align}
\end{widetext}

Here $\rho_c$ is the central density while $c_{s, r}$ and $c_{s,t}$ are the speed of sound in the radial and tangential directions, defined by
\begin{align}
    c_{s,r}^2=\left(\frac{\partial p_r}{\partial \rho}\right)_{\mathrm{eq}},\quad
    c_{s,t}^2&=\left(\frac{\partial p_t}{\partial \rho}\right)_{\mathrm{eq}}\label{Eq: sound speed},
\end{align}
where the subscript ``eq'' denotes the derivative of the background $p_r$ against $\rho$.

Some of these requirements are naturally satisfied for a NS. They include: 
\begin{align}
    \rho\geq0, \,\,\,\rho+p_r\geq0, \,\,\,p_r\geq0,\,\,\,c_{s,r}^2\geq0,
\end{align}
and, if $p_t\geq0$ is satisfied, $\rho+p_t\geq 0$ and $\rho+p_r+2p_t \geq 0$ are also satisfied. 

Let us further reduce some redundant constraints. 
Consider a star that has $p_t<0$ at a certain point inside the star. At the star surface, the radial and tangential pressure must be zero. Therefore, there must exist a point at which the negative tangential pressure increases (or decreases in magnitude) as the radius increases. Since our star generally has $d\rho/dr<0$, this implies that there exists a point at which $c_{s,t}^2=\partial p_t/\partial \rho<0$. Therefore, a violation of the positive tangential pressure condition must also violate the tangential causality condition.

As a result, the criteria that we are left with are
\begin{align}
    &\frac{dM}{d\rho_c}\geq0,\\
    &\rho\geq|p_r|, \,\,\,\rho\geq|p_t|,\\
    &1\geq  c_{s,r}^2, c_{s,t}^2\geq 0.
\end{align}

The maximum-mass criterion holds for anisotropic stars using the modified H-model, i.e. stars such that $M<M_{\rm max}$ are radially stable~\cite{Guedes:2025gqi}. The dominant energy condition generally holds for the range of anisotropy, EOS, and central density that we are studying, except for very few models. Radial stability and tangential causality constraints on the NS mass are shown in Fig.~\ref{fig:placeholder}. For a fixed EOS and anisotropy, the allowed mass range corresponds to the region below the curve that gives the smaller maximum mass limit.

\subsection{$w$-modes}
\label{sec:$w$-modes}
Having spherically-symmetric configurations explained in Sec.~\ref{sec:SS} as background, we next study stellar perturbations to find $w$-modes for anisotropic NSs.

\subsubsection{Numerical Procedures}

We need to solve an eigenvalue problem to find the $w$-mode oscillation frequency $\omega$. The eigenvector $\vec{Y}$ in Eq.~\eqref{eq:eigen_vec} has an arbitrary magnitude as it satisfies homogeneous equations, and we care only about its direction. 
We can solve this vector function with initial conditions from the center of the star and boundary conditions at the surface of the star. At the center of the star, we require the following regularity condition:
\begin{eqnarray}
    \left(\frac{d\vec{Y}}{dr}\right)_{r=0}=0.\label{Eq:eigenvector flatness}
\end{eqnarray}
This gives the relation of initial values at $r=0$: 
\begin{align}
    H_1(0)&=\frac{16\pi}{\ell(\ell+1)} (\rho_0+p_0) W(0)+\frac{2}{l(\ell+1)}K(0) ,\nonumber\\
    X(0)&= e^{\nu_0/2}(\rho_0+p_0) \left(\frac{4\pi}{3} \rho_0+4\pi p_0-\frac{\omega^2}{e^{\nu_0}\ell}+\frac{K(0)}{2}\right),
\end{align}
where
\begin{align}
    K(0)&=\pm (\rho_0+p_0),\nonumber \\
    W(0)&=1,
\end{align}
give the two linearly independent solutions.

The surface boundary condition requires the Lagrangian perturbation of the pressure to vanish. Since $X$ is proportional to the Lagrangian perturbation of the radial pressure, as in Eq. (A7) in~\cite{nonradial}, we choose three independent boundary conditions at the stellar radius $R$: 
\begin{eqnarray}
    \vec{Y}(R)=(H_1(R), K(R), W(R), X(R))\nonumber\\=(1,0,0,0), (0,1,0,0),(0,0,1,0).
\end{eqnarray}
Note that the choices of the numerical values in $\vec{Y}(R)$ are arbitrary as long as they are linearly independent and satisfy $X(R) = 0$. The actual dimensions of the perturbation variables are carried by the coefficients in Eq.~\eqref{eq:a_coefficients}. We solve the perturbation equations numerically from the center to $0.5R$ using two independent vectors, labeled $(\vec{v_1}, \vec{v_2}) $, and we also solve backward from the surface to $0.5R$ using three independent vectors, labeled $(\vec{v_3}, \vec{v_4},\vec{v_5})$. We match the five solutions at $0.5R$ by
\begin{eqnarray}
    a_1 \vec{v_1}+a_2\vec{v_2}=a_3\vec{v_3}+a_4\vec{v_4}+\vec{v_5},\label{eq:a_coefficients}
\end{eqnarray}
to determine the coefficients $a_i$. This would give us the solution to the perturbation equations throughout the star normalized by $\vec{v_5}$. From the perturbation vector at the surface of the star, we can follow Leaver's method~\cite{1985Leaver} to solve the quasi-normal mode frequencies.

\subsubsection{Results}

\begin{figure}
    \centering
    \includegraphics[width=1\linewidth]{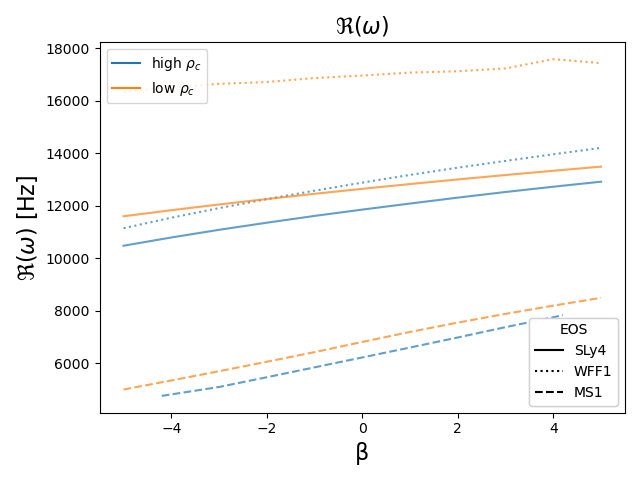}
    \includegraphics[width=1\linewidth]{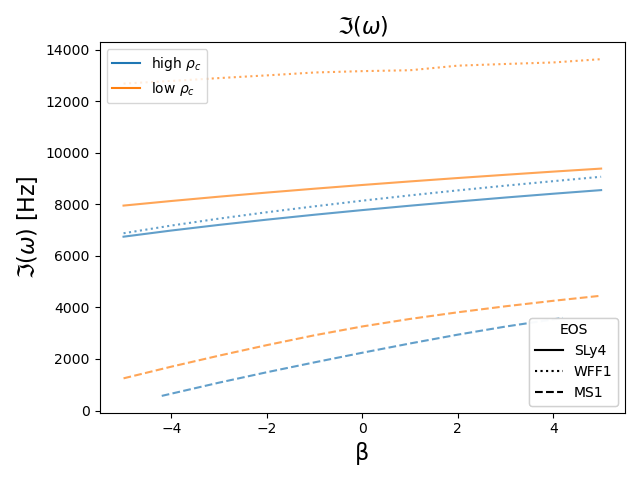}
    \caption{ The real (top) and imaginary (bottom) parts of the scaled frequency for $w$-modes as a function of the anisotropy parameter $\beta$ for various EOSs with the central energy density of $\rho_c = 1\times10^{15}$~g/cm$^3$ (high $\rho_c$) and $\rho_c = 8\times10^{14}$~g/cm$^3$ (low $\rho_c$). For MS1 EOS, the high central density star ceases to be physical for $|\beta| >4.2$.}
    \label{fig:freqpresent}
\end{figure}

We now present our numerical findings on the $w$-modes for anisotropic NSs. Figure~\ref{fig:freqpresent} presents the real and imaginary parts of the $w$-mode frequency against the anisotropy parameter $\beta$ for three different EOSs with two different central energy densities $\rho_c$. Observe that both the real and imaginary frequencies increase as one increases the anisotropy. Similarly to the isotropic case, the real and imaginary parts of the frequencies are higher for softer EOSs and lower $\rho_c$.

\begin{figure}
    \centering
    \includegraphics[width=1\linewidth]{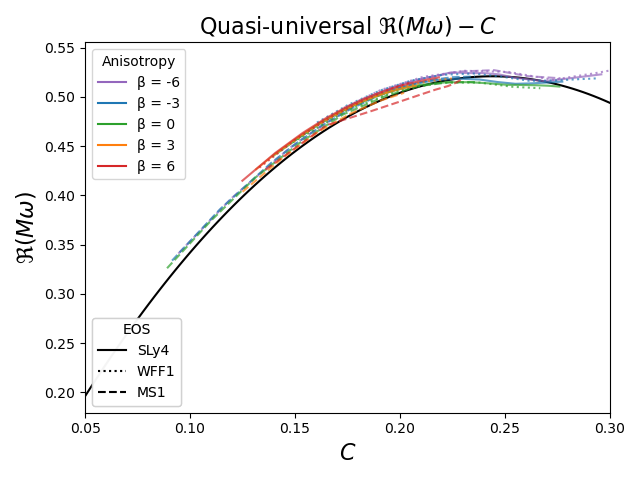}
    \includegraphics[width=1\linewidth]{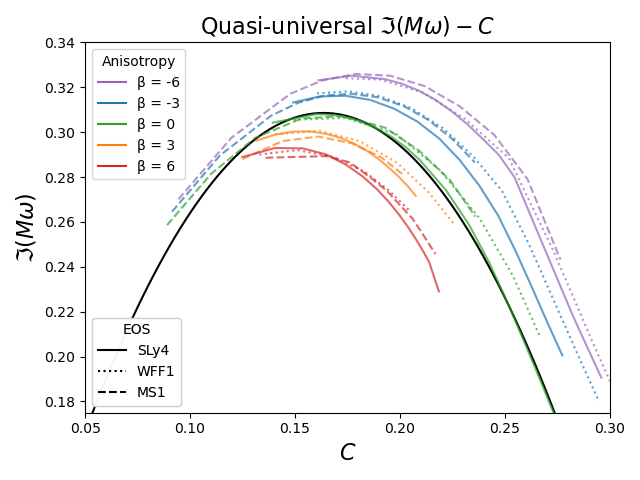}
    \caption{The real (top) and imaginary (bottom) parts of the scaled frequency for $w$-modes as a function of compactness, $C$, for anisotropy parameters and EOSs. The black curve, denoted as TL, is the fit for isotropic NSs found by Tsui and Leung~\cite{TL} using the Tolman VII model. Observe that the relation for the real frequency remains quasi-universal to variation in both EOSs and anisotropy parameters, while the one for the imaginary frequency remains to be quasi-universal for a fixed anisotropy.}
    \label{fig:Universal}
\end{figure}

\begin{figure}
    \centering
    \includegraphics[width=0.99\linewidth]{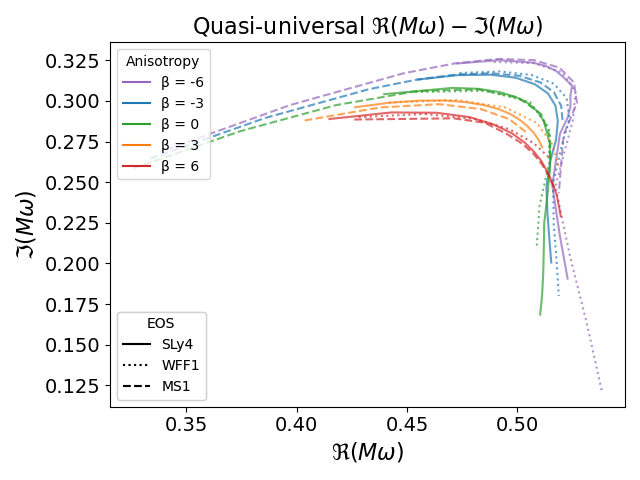}\
    \caption{Relation between the real and imaginary parts of the scaled frequency for $w$-modes with various anisotropy parameters and EOSs.
}
    \label{fig:Universal_2}
\end{figure}

The $w$-mode frequencies have been found to enjoy quasi-universal relations for isotropic NSs~\cite{Benhar:2004xg,TL,Tsui:2005zf}. Let us now study whether such universality holds even for anisotropic NSs. Figure~\ref{fig:Universal} presents the real and imaginary parts of the $w$-mode scaled frequencies, $M\omega$, against the stellar compactness $C = M/R$ for various EOSs and anisotropy parameters. We also show the fit in~\cite{TL} for isotropic NSs. For the real frequency, observe that the relation found for isotropic NSs is approximately valid even for anisotropic NSs. Namely, the relation is quasi-universal to variation in both EOSs and anisotropy. On the other hand, the relations with the imaginary frequency have a clear anisotropy dependence. Interestingly, the relations remain insensitive to the EOS (i.e. $p_r-\rho$ relations) for fixed anisotropy. The latter is similar to the quasi-universal relation between the fundamental mode frequency and the tidal deformability for anisotropic NSs found in~\cite{Guedes:2025gqi}. One can further eliminate the compactness dependence to find the relation between the real and imaginary parts of the $w$-mode frequencies, as shown in Fig.~\ref{fig:Universal_2}. Similar to the relation between the imaginary frequency and compactness, the relations have a significant anisotropy dependence, while they remain EOS-insensitive for fixed anisotropy. 

We next study the stability of $w$-modes for anisotropic NSs, which can be checked from the sign of $\Im (\omega)$. Namely, the mode becomes stable (unstable) if $\Im(\omega)$ is positive (negative). Within the physically-viable parameter space for our anisotropic NS models, we did not find any unstable $w$-modes. On the other hand, we identified unstable $w$-modes once we relax some of the physically-viable conditions, i.e. the causality on the tangential sound speed for the specific parameters we chose. 

Figure~\ref{fig:showunstable} presents the $w$-mode frequency against the anisotropy parameter $\beta$ for NSs with MS1 EOS and the central energy density of $\rho_c=6\times 10^{14}$g/cm$^3$. For $\beta \lesssim -8$, NSs are not physically viable as the tangential sound speed can exceed unity. For the above EOS and central density, we found that $\Im(\omega)$ can go negative when $\beta \lesssim -13.2$, indicating that $w$-modes become unstable for such unphysical NSs with largely-negative (i.e.~tangential pressure dominated) anisotropy. 

\section{Conclusions and Discussions}
\label{sec:conclusion}
In this work, we computed, for the first time, $w$-mode oscillations for anisotropic NSs following the perturbation framework in full general relativity developed in~\cite{nonradial} by some of the authors. For the phenomenological anisotropy model used in~\cite{nonradial,Guedes:2025gqi} that is an extension of the one proposed by Horvat \textit{et al}.~\cite{Horvat_2011}, we found the real and imaginary parts of the $w$-mode frequencies to increase as the anisotropy parameter $\beta$ increases for a fixed central density. We found the relation between $\Re(M \omega)$and $C$ to be insensitive to both EOS and the anisotropy $\beta$. On the other hand, the one between $\Im(M \omega)$ and $C$ depends sensitively on $\beta$ while it remains almost EOS-universal for a fixed $\beta$. We did not identify any unstable $w$-modes within the physically-viable parameter space, while the modes can be unstable for largely-negative $\beta$ when we consider unphysical parameter regions. In~\cite{nonradial}, we found that $p$-modes can be unstable for anisotropic stars with the same phenomenological model as considered here, and provided an analytic justification on why the modes can be unstable for non-vanishing anisotropy through a variation method following~\cite{ipser1973} for isotropic NSs. Such an analysis applies to all $\ell \geq 2$ polar modes, and hence, our finding of unstable $w$-modes for anisotropic NSs is consistent with the analysis in~\cite{nonradial}. 

\begin{figure}
    \centering
    \includegraphics[width=1\linewidth]{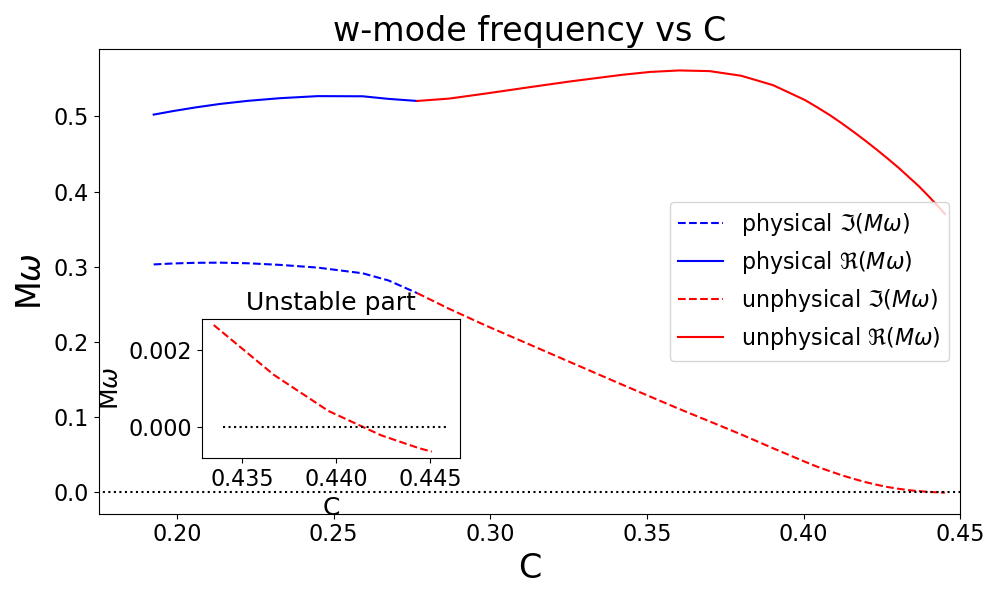}
    \caption{
    Similar to Fig.~\ref{fig:showunstable}, we present the unstable mode for MS1 EOS and a fixed central density of $6\times 10^{14}$ g/cm$^3$ as a function of compactness $C$.}
    \label{fig:unstable compare}
\end{figure}

There is a possibility that the unstable mode we found may not correspond to $w$-modes in our original classification scheme at less negative $\beta$, as the mode may have encountered an avoided crossing with other modes when we decrease $\beta$. The sharp change in slope of the real frequencies in Figs.~\ref{fig:showunstable} and \ref{fig:unstable compare} may be due to a close encounter of the $w$-mode with another quasi-normal mode. In classical stellar pulsation theory, the oscillation frequency of each mode can be viewed as a smooth function of a varying parameter (e.g. the anisotropy parameter $\beta$ or the stellar compactness $C$). When the parameter varies, the eigenfrequencies of these modes may approach each other.  
If the modes are coupled with each other, or equivalently, they do not originate from completely decoupled wave equations, the mode frequencies will repel each other to avoid a crossing.
The mode characteristics will be exchanged as the eigenfrequencies of the two modes move away from the point of closest approach. This feature is well-known in Hermitian eigenvalue problems and has been identified in classical stellar pulsations (see, e.g.,~\cite{1977A&A}). In non-Hermitian systems, the situation is more complicated as there can be avoided crossings in real frequency and true crossings in the imaginary frequency, or the other way round \cite{Rotter_2009_arxiv, Rotter_2009}. Further studies are required to uncover the mode-coupling properties of the $w$-mode with the other modes. That requires a thorough search of all quasi-normal modes at nearby frequencies, which we shall leave as future work. Here, we simply assume an exchange in mode characteristics after an avoided crossing occurs for the real frequency, similar to the case found between the $f$-mode and shear modes in a solid relativistic star \cite{Lau2017thesis}.
Figure~\ref{fig:unstable compare} shows the rescaled $w$-mode frequency as a function of the compactness. Observe that the real part of the frequency does not seem to connect smoothly from the physical branch to the unphysical branch as the compactness increases, which may indicate the presence of avoided crossing. Thus, the unstable mode we have discovered in the unphysical branch may be the result of the avoidance of crossing from another mode.

Previous works have explored the possibility of reducing the complexity of $w$-modes through approximations. In Appendix~\ref{sec:ICA}, we discuss two approximations of $w$-modes, inverse Cowling approximation (ICA)~\cite{Andersson_1996} and generalized inverse Cowling approximation (GICA)~\cite{Wu_2007}. We also discussed the validity of ICA and GICA for anisotropic NSs. 

More work is needed to arrive at more robust conclusions on mode crossing and the validity of GICA on $w$-modes for anisotropic NSs. It would also be important to study anisotropic models other than the one considered here, in particular, for models that are more physically motivated, such as the ones for elastic NSs~\cite{Dong:2024lte} or those modeled based on liquid crystal~\cite{Cadogan:2024mcl,Cadogan:2024ohj,Cadogan:2024ywc}. We leave these studies for future work.

    \begin{acknowledgments}
       V.G., S.A. and K.Y.  acknowledge support from the Owens Family Foundation. S.A. and K.Y. also acknowledge support from NSF Grant PHY-2309066 and PHY-2339969.  S.Y.L. acknowledges support from Montana NASA EPSCoR Research Infrastructure Development under award No. 80NSSC22M0042.

    \end{acknowledgments}

\appendix

\section{Perturbation Equations}
\label{app:pert}

In this appendix, we provide perturbation equations following~\cite{nonradial}. The main equations are given by
\begin{widetext}
\allowdisplaybreaks
\begin{align}
H_1' &= \Bigl[4\pi(\rho - p_r)e^\lambda r
              - \frac{2me^\lambda(\ell+1)}{r^3}\Bigr] H_1  + \frac{e^\lambda}{r}K
              + \frac{e^\lambda}{r}H_0  - \frac{16\pi(\rho + p_r)e^{2\lambda}}{r}(1-\bar\sigma)V,
              \label{Eq:perturbation H1}\\[0.5ex]
K'   &= \frac{\ell(\ell+1)}{2r}H_1
        + \Bigl(\frac{\nu'}{2} - \frac{\ell+1}{r}\Bigr)K  - 8\pi(\rho + p_r)e^{\lambda/2}W
              + \frac{1}{r}H_0,
              \label{Eq:perturbation K}\\[0.5ex]
W'   &= r e^{\lambda/2}(1-\bar\sigma)K
        + \Bigl(-\frac{\ell+1}{r} + \frac{2\bar\sigma}{r}\Bigr)W + \frac{r e^{(\lambda-\nu)/2}}{\gamma p_r}X
              + \frac{r e^{\lambda/2}}{2}H_0 - \frac{\ell(\ell+1)e^{\lambda/2}}{r}(1-\bar\sigma)V,
              \label{Eq:perturbation W}\\[0.5ex]
X'   &= \frac{\rho + p_r}{2}e^{\nu/2}\Bigl[r\omega^2e^{-\nu}
         + \frac{\ell(\ell+1)}{2r}(1-2\bar\sigma)\Bigr]H_1  + \frac{\rho + p_r}{2}e^{\nu/2}
        \Bigl[\Bigl(\tfrac{3}{2}-2\bar\sigma\Bigr)\nu'
               - (1-6\bar\sigma)\frac{1}{r}
               - \frac{4\bar\sigma^2}{r}\Bigr]K \nonumber\\
     &\quad - \frac{\rho + p_r}{r}e^{(\lambda+\nu)/2}
        \bigl[4\pi(\rho + p_r)+ \omega^2e^{-\nu} - F\bigr]W  - \frac{1}{r}
        \Bigl(\ell - 2\frac{\rho + p_r}{\gamma\,p_r}\bar\sigma\Bigr)X \nonumber\\
     &\quad + \frac{\rho + p_r}{2}e^{\nu/2}
        \Bigl(\frac{1}{r} - \frac{\nu'}{2}\Bigr)H_0 + \frac{\ell(\ell+1)e^{\nu/2}}{r^2\,p_r'}(1-\bar\sigma)\,V
              + \frac{2e^{\nu/2}}{r}S.
              \label{Eq:Perturbation X}
\end{align}
The overbarred pressure anisotropy, $\bar{\sigma}$, is defined as
\begin{equation}
    \bar{\sigma}=\frac{\sigma}{p_r+\rho},
\end{equation}
while the adiabatic index, $\gamma$, is defined as
\begin{equation}
    \gamma \equiv 
    \frac{\rho+p_r}{p_r}\left(\frac{\partial p_r}{\partial \rho}\right)_{\mathrm{eq}},
\end{equation}
where the subscript ``eq'' denotes the derivative of the background $p_r$ against $\rho$. $X, F,$ and $S$ are defined as
\begin{align}
    \sum_{\ell,m} r^\ell X Y_{\ell m} e^{i \omega t} =& -e^{\nu/2} \Delta p, \\
    F =& e^{-\tfrac{\lambda}{2}}\Biggl\{\frac{r^2}{2}\biggl(e^{-\tfrac{\lambda}{2}}\frac{\nu'}{r^2}\biggr)'-e^{-\tfrac{\lambda}{2}}\Bigl[\Bigl(\frac{6}{r^2}-\frac{2\nu'}{r}\Bigr)\bar\sigma 
-\frac{2}{r}\Bigl(\frac{r\sigma'}{\rho+p_r}\Bigr)-\frac{4\,\bar\sigma^2}{r^2}\Bigr]\Biggr\}.\\
S =& - \left[ \left( \frac{\partial \sigma}{\partial p_r} \right)_{\text{eq}} + (\rho + p_r) \left( \frac{A}{p'_r} + \frac{1}{\gamma p_r} \right) \left( \frac{\partial \sigma}{\partial \rho} \right)_{\text{eq}} \right] e^{-\lambda/2 }p'_r \frac{W}{r} \nonumber \\
&- \left[ \left( \frac{\partial \sigma}{\partial p_r} \right)_{\text{eq}} + \frac{\rho + p_r}{\gamma p_r} \left( \frac{\partial \sigma}{\partial \rho} \right)_{\text{eq}} \right] e^{-\nu/2} X - \left( \frac{\partial \sigma}{\partial \mu} \right)_{\text{eq}} e^{-\lambda} H_0.
\end{align}
The Schwarzschild discriminant is defined as
\begin{eqnarray}
    A=\frac{\rho'}{\rho+p_r}-\frac{p_r'}{\gamma p_r}=0,
\end{eqnarray}
and is taken to be zero, assuming the EOS is a single variable function of $p_r$. 
There are also algebraic relations connecting perturbation variables:
\begin{align}
\left[3m + \frac{(\ell-1)(\ell+2)}{2}\,r + 4\pi r^3 p_r\right]H_0=&8\pi r^3 e^{-\nu/2}X-\left[\frac{\ell(\ell+1)}{2}(m+4\pi r^3 p_r)- \omega^2 r^3 e^{-\lambda-\nu}\right]H_1\nonumber\\ 
&+ \left[\frac{(\ell-1)(\ell+2)}{2}\,r
      - \omega^2 r^3 e^{-\nu}
      - \frac{e^{-\lambda}}{r}(m+4\pi r^3 p_r)(3m - r + 4\pi r^3 p_r)\right]\,K\nonumber\\ 
&- 16\pi\,r\,e^{-\lambda/2}(\rho+p_r)\, W \\
X
=& \omega^2(\rho+p_r)\,e^{-\nu/2}(1-\bar\sigma)\,V
+ \frac{\rho+p_r}{2}\,e^{\nu/2}\,H_0
- \frac{p_r'}{r}\,e^{-(\lambda+\nu)/2}\,W
- e^{\nu/2}\,S\sid{.}\label{Eq: X}
\end{align}
\end{widetext}

\section{(Generalized) Inverse Cowling Approximation}
\label{sec:ICA}

\begin{figure}
    \centering
    \includegraphics[width=0.99\linewidth]{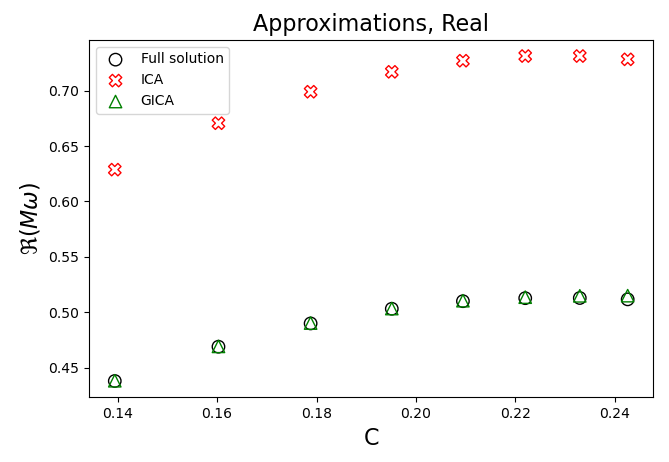}
    \includegraphics[width=0.99\linewidth]{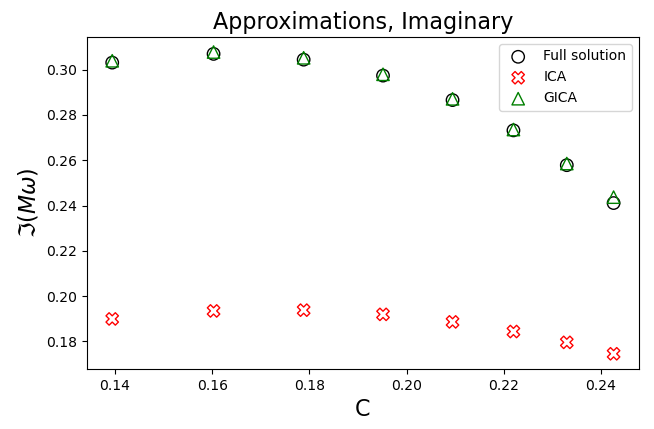}
    \caption{Real (top) and imaginary (bottom) parts of $w$-mode frequencies against the stellar compactness for \emph{isotropic} NSs with SLy4 EOS. We present results for the full calculation, as well as approximation solutions within ICA and GICA. Notice that the results for ICA are off from the full ones, while GICA can accurately reproduce the full results.}
    \label{fig:ICA}
\end{figure}

In this appendix, we provide the validity of ICA and GICA for \emph{isotropic} NSs.
One possible way to show the stability of the $w$-modes is to reduce the complexity of the system through valid approximations. The Cowling approximation~\cite{Cowling:1941nqk} has been used to compute fluid modes (like $f$-modes and $p$-modes) by ignoring spacetime perturbations. $w$-modes originate from spacetime perturbation, and hence, one would expect minimal involvement of matter perturbation,  a key distinction from other fluid oscillation modes. For $w$-modes, ICA was first adopted, where one keeps only spacetime perturbations while ignoring fluid displacements $V$ and $W$ as well as the Lagrangian perturbation to radial pressure $X$~\cite{Andersson_1996}. Although ICA includes the essential features of $w$-modes, the numerical values of the scaled frequency under ICA significantly disagree with exact solutions (see Fig.~\ref{fig:ICA} for the isotropic NS case).

\begin{figure}
    \centering
    \includegraphics[width=0.99\linewidth]{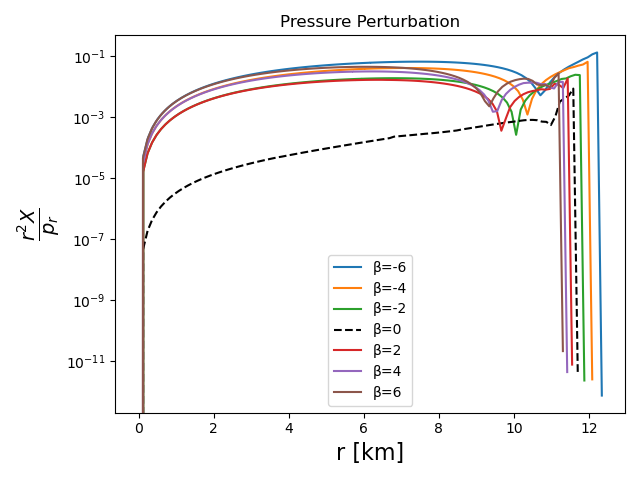}
    \caption{ The profile for the ratio between the $\ell=2$ Lagrangian pressure perturbation $r^2X$ and background radial pressure $p_r$  for the central density of $10^{15}$g/cm$^3$ with SLy4 EOS and various anisotropy parameter $\beta$. We see that for $\beta=0$, $r^2X/p_r$ is highly suppressed, justifying GICA, while this is no longer the case for $\beta \neq 0$.}
    \label{fig:Xpratio} 
\end{figure}
By observing the Lagrangian perturbation of pressure to be negligible for $w$-modes, Wu and Leung~\cite{Wu_2007} proposed GICA of spacetime modes. The Lagrangian perturbation of the pressure is described by the physical quantity $X$. In GICA, both $X(r)$ and $X'(r)$ are taken to be zero, while keeping $V$ and $W$. Eventually, this reduces the 4-dimensional eigenvalue problem to one of  2-dimensions. Reference~\cite{Wu_2007} has shown GICA to be effective and strongly aligned with full solutions of $w$-modes for isotropic NSs (once again, see Fig.~\ref{fig:ICA}). 

We comment on the potential validity of these approximations on anisotropic NSs. First, with ICA, all of the anisotropy dependences at the level of the perturbations are forced to vanish (except for the ones originating from the anisotropy dependence in background solutions), and hence the approximation is expected to be invalid (which was already the case even for isotropic NSs). To infer the validity of GICA for anisotropic NSs, Fig.~\ref{fig:Xpratio} presents the profile of the ratio $r^2X/p_r$ inside a star of fixed central density and EOS but with varied anisotropy parameter $\beta$. This dimensionless ratio signifies the amount of pressure perturbation when $\ell=2$ oscillations are excited. We see that $r^2X/p_r$ is highly suppressed for the isotropic case with $\beta =0$, while the ratio becomes significantly higher for $\beta \neq 0$. 

This seems to be inconsistent with the fundamental assumption of GICA, and thus we would expect GICA to be invalid for $w$-modes with anisotropic pressure.
\newpage

\bibliography{w-mode}

\end{document}